\documentclass[preprint, amsmath,amssymb, aps]{revtex4-1}
\usepackage{dcolumn}
\usepackage{amssymb}
\usepackage{subfig}
\usepackage{caption}
\usepackage{float}
\usepackage{graphicx}
\usepackage{epsfig}
\usepackage{amsmath}
\usepackage{float}
\usepackage{epsfig}
\usepackage{appendix}
\begin{document}

\title{Rate constant calculations for diffusion in presence of a Gaussian sink: An exact analytical solution}
\author{Swati Mudra* and Aniruddha Chakraborty}
\affiliation{School of Basic Sciences, Indian Institute of Technology Mandi, Kamand, HP-175075, India.}
\date{\today}
\begin{abstract}
\noindent In this paper, we have developed a method to solve Smoluchowski equation in presence of a Gaussian sink. We have solved the one dimensional  equation for a flat potential. Our method provides solution in Laplace domain, which is used to derive an analytical expression for time average rate constant. Our solution can be used to analyze several related problems involving diffusion-reaction systems.
\end{abstract}

\maketitle
\section{Introduction}
\noindent Diffusion-reaction system has a long history as an important topic of both theoretical and experimental research \cite{Risken,Nishijima,Agmon,Dobler,Mathies,Hong,Archer,Luczkaa,Ansari,Szabo,Oster,Dagdug}. The most commonly used model of diffusion-reaction system use the one dimensional Smoluchowski equation  modified by the addition of a sink term \cite{Nishijima,Agmon,Dobler,Mathies,Hong,Archer,Luczkaa,Ansari,Szabo,Oster,Dagdug}. Different varieties of sink functions are used to model different problems. Among those sink functions, the most popular ones are, pinhole sink, gaussian sink \cite{Bagchi2,Berezhkovskii}, lorentzian \cite{Bagchi}, Dirac delta function sink \cite{Sebastian1,Sebastian2,Chakravarti}, and exponential sink \cite{Rajarshi}. Among all these sinks, only the case with  Dirac delta function sink has analytical solution \cite{Sebastian1,Bagchi1}. In the following, we derive the exact analytical solution for the case with Smoluchowski equation for flat potential with an gaussian sink of arbitrary strength. In our model the sink is located at origin. Our model is very much suitable for understanding the problems like electronic relaxation of a molecule in polar solvent, electron transfer in chemical and biological reactions, diffusion through cell membrane etc. Here we will discuss the problem of electronic relaxation in solution in brief. Molecular configuration of a molecule in polar solvent does random walk on both the ground and excited electronic state potential energy surface. Molecule can be placed on the electronically excited state potential energy surface by the application of light. The electronically excited  molecule can get relaxed by radiative decay process and by non-radiative decay process \cite{Bagchi}. It is known that the rate of radiative decay is same from all the molecular configurations \cite{Bagchi} and the non-radiative decay is possible only from certain specified molecular configurations \cite{Oster}.

\par
As we have discussed that there are many sink models has been introduced to understand the process. Gaussian sink model is one of the important and very difficult to solve analytically. In this work we will present a new methodology to solve smoluchowski equation for a flat potential in presence of a gaussian sink. Further we will use this solution to calculate the average rate constant for a harmonic potential case. Here we consider the case where non-radiative decay is the only possible way of electronic relaxation. One dimensional Smoluchowski equation is used to model the random change of the molecular configurations and the non-radiative decay process is modeled by adding a molecular configuration dependent sink term to the same equation. In the following we use the variable $x$ to denote molecular configurations, therefore the problem can be thought of as a particle diffusing in flat potential in presence of a sink function and the corresponding probability distribution obey the following modified Smoluchowski equation as given below
\begin{equation}
\frac{\partial P(\xi,\tau)}{\partial \tau} = \left(D \frac{\partial^2}{\partial \xi^2} - k_0 e^{-\alpha(\xi-\xi_c)^2} \right)  P(\xi,\tau),
\label{eqn:Gaus-1}
\end{equation}
\noindent where $P(\xi,\tau)$ is the probability that the particle can be found at $\xi$ at time $\tau$, $k_0$ represent the non-radiative decay rate constant, $\xi_c$ is the peak of the sink and $D$ represents the diffusion constant.

\section{Solution of Smoluchowski equation for a special harmonic potential}
\noindent In the following, we will solve this problem for a specific harmonic potential case where the force constant $k$ will be taken as $\alpha D$. So for convenience we will start with the following equation
\begin{equation}
\frac{\partial P(x,t)}{\partial t} = D \frac{\partial^2}{\partial x^2} P(x,t) +\alpha D \frac{\partial}{\partial x}[xP(x,t)] - k_0 S(x) P(x,t).
\label{eqn:Gaus-3}
\end{equation}
\noindent We will solve this equation for a Gaussian sink function {\it i.e.}, $S(x) = e^{-\alpha x^2}$ which is located at the origin and we will derive an analytical expression for survival probability in the Laplace domain. Laplace transformation of Eq.(\ref{eqn:Gaus-3}) for a Gaussian sink yields the following equation
\begin{equation}
s\tilde P(x,s)-P(x,0)=D \frac{\partial^2\tilde P(x,s)}{\partial x^2} +\alpha D x  \frac{\partial {\tilde P}(x,  {\tilde s})}{\partial x}+\alpha D P(x,s) - k_0 e^{-\alpha x^2} \tilde P(x,s).
\label{eqn:Gaus-4}
\end{equation}
\noindent The initial probability distribution on the electronically excited state $P(x,0)$, can be assumed to be the same as it was in the ground electronic state just before excitation (with the Franck-Condon assumption) using electromagnetic radiation of appropriate frequency. Therefore, Eq.(\ref{eqn:Gaus-4}) is now modified as
\begin{equation}
 (s-\alpha D) {\tilde P}(x, {\tilde s}) - \delta(x-x_0)  =  D \frac {\partial^2 {\tilde P}(x,  {\tilde s})}{\partial x^2} +\alpha D x  \frac{\partial {\tilde P}(x,{\tilde s})}{\partial x} - k_{0} e^{ - \alpha x^2} {\tilde P}(x,{\tilde s}).
 \label{eqn:Gaus-5}
\end{equation}
\noindent We will solve the above equation for $s=\alpha D$, then we will get
\begin{equation}
 - \delta(x-x_0)  =  D \frac{\partial^2 {\tilde P}(x,\alpha D )}{\partial x^2} +\alpha D x \frac{\partial {\tilde P}(x, \alpha D )}{\partial x} - k_{0} e^{ - \alpha x^2} {\tilde P}(x,\alpha D ).
 \label{eqn:Gaus-6}
\end{equation}
\noindent Now first we consider the case, where $x \neq x_0$, therefore the above equation becomes
\begin{equation}
0 =  D \frac{\partial^2 {\tilde P}(x,\alpha D)}{\partial x^2}+\alpha D x  \frac{\partial {\tilde P}(x,\alpha D)}{\partial x} - k_{0} e^{ - \alpha x^2} {\tilde P}(x,\alpha D).
 \label{eqn:Gaus-7}
\end{equation}
\noindent Now we can re-write the above equation as 
\begin{equation}
0 =  e^{ \alpha x^2} \frac{\partial^2 {\tilde P}(x, \alpha D)}{\partial x^2} + \alpha x e^{ \alpha x^2} \frac{\partial {\tilde P}(x,\alpha D)}{\partial x}- \frac{k_{0}}{D} {\tilde P}(x,\alpha D).
 \label{eqn:Gaus-8}
\end{equation}
\noindent Again 
\begin{equation}
0 =  \left[e^{ \alpha x^2/2} \frac{\partial}{\partial x} + \sqrt{\frac{k_{0}}{D}}\right] \left[e^{ \alpha x^2/2} \frac{\partial}{\partial x} - \sqrt{\frac{k_{0}}{D}}\right] {\tilde P}(x,\alpha D).
\label{eqn:Gaus-9}
\end{equation}
\noindent or  
\begin{equation}
0 =  \left[e^{ \alpha x^2/2} \frac{\partial}{\partial x} - \sqrt{\frac{k_{0}}{D}}\right] \left[e^{ \alpha x^2/2} \frac{\partial}{\partial x} + \sqrt{\frac{k_{0}}{D}}\right] {\tilde P}(x,\alpha D).
\label{eqn:Gaus-10}
\end{equation}
\noindent Therefore we have either 
\begin{equation}
0 =   \left[e^{ \alpha x^2/2} \frac{\partial}{\partial x} + \sqrt{\frac{k_{0}}{D}}\right] {\tilde P}(x,\alpha D).
\label{eqn:Gaus-11}
\end{equation} 

\noindent or
\begin{equation}
0 =   \left[e^{ \alpha x^2/2} \frac{\partial}{\partial x} - \sqrt{\frac{k_{0}}{D}}\right] {\tilde P}(x,\alpha D).
\label{eqn:Gaus-12}
\end{equation} 
\noindent The solution of Eq.(\ref{eqn:Gaus-11}) is given by
\begin{equation}
{\tilde P}(x,\alpha D) = A e^{ -\frac{\sqrt{\pi } \sqrt{\frac{k_0}{D}} Erf\left(\sqrt{\frac{\alpha}{2}} x \right)}{\sqrt{2 \alpha }}}.
\label{eqn:Gaus-13}
\end{equation}
\noindent The solution of Eq.(\ref{eqn:Gaus-12}) is given by
\begin{equation}
{\tilde P}(x,\alpha D) = A e^{ \frac{\sqrt{\pi } \sqrt{\frac{k_0}{D}} Erf\left(\sqrt{\frac{\alpha}{2}} x \right)}{\sqrt{2 \alpha }}}.
\label{eqn:Gaus-14}
\end{equation}
\noindent Eq.(\ref{eqn:Gaus-13}) and (\ref{eqn:Gaus-14}) both are the solutions of Eq.(\ref{eqn:Gaus-8}) for $x > x_0$ and $x <x_0$. The unknown constant $A$ can be determined by using the following boundary condition in Eq.(\ref{eqn:Gaus-6})
\begin{equation}
 - 1  =  D \left[\frac{\partial}{\partial x} \left( {\tilde P}(x,\alpha D - k_r)\right)\right]^{x={x_0}+\epsilon}_{x={x_0} - \epsilon}.
\label{eqn:Gaus-15}
\end{equation}
\noindent From here we can calculate the constant A which is as follows
\begin{equation}
 A= \frac{e^{\alpha {x_0}^2 /2}}{\sqrt{D k_0}\left[e^{- \frac{\sqrt{\pi } \sqrt{\frac{k_0}{D}} Erf\left(\sqrt{\frac{\alpha}{2}} x_0 \right)}{\sqrt{2 \alpha }}}+e^{ \frac{\sqrt{\pi } \sqrt{\frac{k_0}{D}} Erf\left(\sqrt{\frac{\alpha}{2}} x_0 \right)}{\sqrt{2 \alpha }}}\right]}.
\label{eqn:Gaus-16}
\end{equation}
\noindent Now we can get the complete solution for $P(x,\alpha D)$ by substituting the value of A into Eq.(\ref{eqn:Gaus-13}) and Eq.(\ref{eqn:Gaus-14}) which is the probability distribution for diffusion on a harmonic potential in presence of a Gaussian sink. Here the force constant has a specific value {\it i.e.} $\alpha D$. Now we use this result to calculate the probability distribution for diffusion on a flat potential in presence of a Gaussian sink.

\section{Rate constant calculations calculations for a flat potential case}
\noindent In this section we will derive the technique to convert the solution for a special harmonic potential into the solution of a flat potential. In this method we have assumed that the gaussian function sink is located at origin with an arbitrary initial condition. Average rate constant in Laplace domain can be defined as
\begin{equation}
 {k_I}^{-1} = {\tilde P}(0).
 \label{eqn:Gaus-17}
\end{equation}
\noindent Here ${\tilde P}(\xi,s)$ is the Laplace transform of $P(\xi,\tau)$. We can find ${\tilde P}(0)$ as
\begin{equation}
{\tilde P}(0) = \int^{\infty}_{-\infty} d\xi \int^{\infty}_{0} P(\xi,\tau)  d\tau .
\label{eqn:Gaus-18}
\end{equation}
\noindent Now using the transformation explained in Appendix we can get
\begin{equation}
{\tilde P}(0) = \int^{\infty}_{-\infty}dx \int^{\infty}_{0} P(x,t) e^{-2kt} dt .
\label{eqn:Gaus-19}
\end{equation}

\noindent Here $P(x,t)$ represents the probability distribution for harmonic potential with force constant $k$. Now let us add a term $- k_r P(x,t)$ in Eq.(\ref{eqn:Gaus-3}), the solution of the equation will modify as
\begin{equation}
{\tilde P}(0)  = \int^{\infty}_{-\infty} dx \int^\infty_{0}P(x,t) e^{-2kt} e^{-k_r t} dt.
\label{eqn:Gaus-20}
\end{equation}
\noindent Now let us take a special case where $k_r = -k$, the above equation will become
\begin{equation}
{\tilde P}(0)  = \int^{\infty}_{-\infty} dx\int^\infty_{0}P(x,t) e^{-kt} dt.
\label{eqn:Gaus-21}
\end{equation}
\noindent which will become
\begin{equation}
{\tilde P}(0)  = \int^\infty_{- \infty} dx {\tilde P}(x,k).
\label{eqn:Gaus-22}
\end{equation}

\noindent In our calculation we have assumed that $k= \alpha D$. So by substituting it in Eq.(\ref{eqn:Gaus-22}) we will get the average rate constant for flat potential with a Gaussian sink by using Eq.(\ref{eqn:Gaus-13}), Eq.(\ref{eqn:Gaus-14}) and Eq.(\ref{eqn:Gaus-16}) as follows
\begin{equation}
{k_I}^{-1}  = \int^\infty_{- \infty} dx {\tilde P}(x,\alpha D).
\label{eqn:Gaus-22}
\end{equation}
\noindent We found $\xi= x- x_0 e^{-kt}$ (see Appendix), similarly $\xi_c= x_c- x_0 e^{-kt}$ from here we can see that $\xi_0 = 0$ for all values of $x_0$. We have taken $x_c =0$ in Eq.(\ref{eqn:Gaus-3}) which makes $\xi= x+ \xi_c$ and for $\xi_0 = 0$, $x_0 =-\xi_c$. In that case $x_0$ will represent the distance between sink and initial position for flat potential. 

\section{Results and Discussion}
\begin{figure}
    \centering
    \includegraphics[scale=0.6]{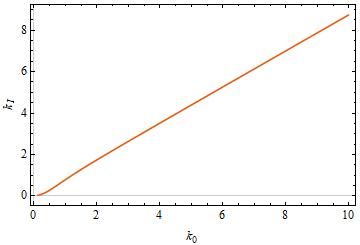}
    \caption{This plot is to understand the behaviour of average rate constant $k_I$ corresponding to decay constant $k_0$. Other parameters in this plot are: $\xi_0=0$, $\xi_c = -1$,, $D=0.5$ and $\alpha= 0.1$.}
    \label{fig:k0-kIgauss}
\end{figure}

\begin{figure}
    \centering
    \includegraphics[scale=0.6]{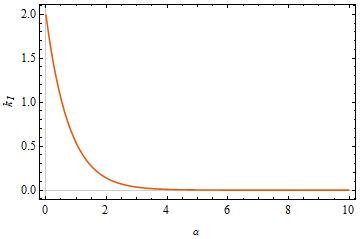}
    \caption{A plot between average rate constant $k_I$ and $\alpha$.Other parameters in this plot are: $\xi_0=0$, $\xi_c = -1$, $D=0.1$ and $k_0 = 2$.}
    \label{fig:alpha-kIgauss}
\end{figure}

\begin{figure}
    \centering
    \includegraphics[scale=0.6]{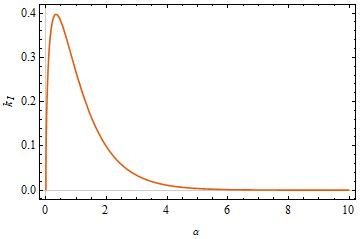}
    \caption{A plot between average rate constant $k_I$ and $\alpha$. Other parameters in this plot are: $\xi_0=0$, $\xi_c = -1$,  $D=0.1$ and $k_0 = 2 \sqrt{\alpha/\pi}$.}
    \label{fig:nor-alpha-kIgauss}
\end{figure}

\begin{figure}
    \centering
    \includegraphics[scale=0.6]{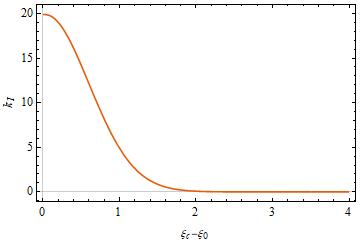}
    \caption{Plot of $k_I$ with respect to the distance between sink position and initial position $\xi_c -\xi_0$. Other parameters in this plot are: $\xi_0=0$,  $\alpha = 1$, $D=0.1$ and $k_0 = 20$.}
    \label{fig:kI-sinkposi}
\end{figure}

\noindent As we can see in section III that we got an exact expression for the average rate constant ${k_I}^{-1}$. This rate constant is an important quantity which can affect the dynamics of the system significantly. In Fig.{\ref{fig:k0-kIgauss}} we can see that $k_I$ is increasing with increasing $k_0$. This is very obvious because $k_0$ is a factor that triggers the reaction. $\alpha$ is a parameter which is associated with the width of sink. Width of the Gaussian will decrease by increasing $\alpha$. Here we will discuss two cases: Fig.{\ref{fig:alpha-kIgauss}} is a plot showing the variation in $k_I$ with respect to $\alpha$ where the sink function is not normalized, so increasing $\alpha$, will narrow down the sink with a constant height and this leads to a decrease in the area of the sink. The plot shows that by decreasing the area of sink the reaction slows down {\it i.e.} the average rate constant decreases. Another case where the sink is a normalised Gaussian function, area will always be the same by changing $\alpha$. In this case the height of Gaussian will increase by increasing $\alpha$. Let us see Fig.{\ref{fig:nor-alpha-kIgauss}}, for $\alpha =0$ the sink term will be zero so no reaction will take place which makes the average rate constant zero. When we start increasing $\alpha$, the reaction starts. Let us expend the sink term {\it i.e.}
\begin{equation}
\sqrt{\alpha/\pi}e^{-\alpha x^2} = \sqrt{\alpha/\pi}\left[1- \alpha x^2 + \frac{\alpha^2 x^4}{2}- \frac{\alpha^2 x^4}{3!}.... \right]
\label{eqn:Gaus-23}
\end{equation}

\noindent for smaller values of $\alpha$, higher order terms will not contribute so sink will be a linear function of $\alpha$. In that case by increasing $\alpha$, sink strength will increase and the average rate constant will also go up. But after sufficiently higher value of $\alpha$ the sink strength will decay exponentially and this decay is more in comparison to $\sqrt{\alpha}$ which leads a fall in the average rate constant. Here we can see that the the role of $\alpha$ is very important in both the cases. It would be interesting to see the effect of the distance between sink position $\xi_c$ and initial position $\xi_0$. Fig.\ref{fig:kI-sinkposi} shows that an increase in $\xi_c -\xi_0$ is making the rate of reaction slower. 

\section{Conclusions}
\noindent Gaussian sink is an important model because it is the most general shape to model the electronic relaxation between two states \cite{Bagchi}. Numerical solution have been provided earlier for this problem \cite{Bagchi,Bagchi1}. Solution of this problem has been reported earlier with some restriction on the width, position and strength of the sink in \cite{Berezhkovskii}. In this paper we have solved the flat potential which has transnational symmetry so the position of the sink is not a very important entity. The real molecular surfaces are harmonic in nature so there is chance of advancement in this area \cite{SA}. In our model sink strength is general and we have solved this problem without any restrictions on sink width so this model is quite applicable to electron transfer and many other biological problems.

\section*{Acknowledgement}
I would like to thank IIT Mandi for HTRA fellowship and resources and another author (A.C.) thanks IIT Mandi for PDA funds.

\section*{Appendix}{}
\renewcommand{\theequation}{A\arabic{equation}}
\setcounter{equation}{0}

Let us take $P(x,t)$ is the solution of Smoluchowski equation {\it i.e.} Eq.(\ref{eqn:Gaus-3}) without the sink term. Which has been reported as
\begin{equation}
P(x, t) = \frac{e^{-\frac{(x-x(t))^2}{4 D \sigma(t)^2}}}{ \sqrt{4 D \pi}\sigma(t)} ,
\label{eqn:A1}
\end{equation}
\noindent here $x(t)=x_0 e^{-kt}$; $\sigma(t)^2 = \frac{1}{2k}(1 - e^{- 2kt})$. Let us use this $P(x,t)$ in Eq.(\ref{eqn:Gaus-3}) without sink term in RHS, we will get
\begin{equation}
e^{2kt}\left[\frac{\partial P(x,t)}{\partial t}-e^{-kt}x_0 k \frac{\partial P(x,t)}{\partial x}\right] = D\frac{\partial^2 P(x,t)}{\partial x^2} P(x,t) 
\label{eqn:A2}
\end{equation}
Now if we take $\xi=x-x_0 e^{-kt}$ and $t=t'$ we will get
\begin{equation}
e^{2kt'}\frac{\partial P(\xi,t')}{\partial t'} = D\frac{\partial^2 P(\xi,t')}{\partial \xi^2}
\label{eqn:A3}
\end{equation}
\noindent Now we will consider
\begin{equation}
\frac{\partial t}{\partial \tau} = e^{ 2 k t'}.
\label{eqn:A4}
\end{equation}
\noindent and we will get
\begin{equation}
\frac{\partial P(\xi,\tau)}{\partial \tau} = D\frac{\partial^2 P(\xi,\tau)}{\partial \xi^2} 
\label{eqn:A5}
\end{equation}
\noindent Eq.(\ref{eqn:A5}) is the smoluchowski equation for flat potential. Where $\tau$ is
\begin{equation}
\tau = \frac{1}{2k}(1-e^{-2kt'}).
\end{equation}
\noindent Here we get that $P(\xi,\tau)$ represents the probability distribution for a flat potential.This method has been described in \cite{Swati} in details.

\end{document}